\documentclass[12pt]{article}
\usepackage{amsmath,amssymb}
\usepackage{a4wide}
\usepackage{color}

\newcommand{\q}{q_\Phi}

\DeclareMathOperator{\tr}{tr}

\begin{document}

\title{
Dynamical Rearrangement of Theta Parameter\\
in Presence of Mixed Chern-Simons Term\bigskip\bigskip\\
}

\author{
Naoyuki {Haba},$^{1}$\footnote{E-mail: \tt haba@phys.sci.osaka-u.ac.jp} \ 
Yoshiharu {Kawamura},$^{2}$\footnote{E-mail: \tt haru@azusa.shinshu-u.ac.jp} \  
and Kin-ya {Oda},$^{1}$\footnote{E-mail: \tt odakin@phys.sci.osaka-u.ac.jp}\bigskip\\
\normalsize$^{1}$\it Department of Physics, Osaka University, Toyonaka 560-0043, Japan\\
\normalsize$^{2}$\it Department of Physics, Shinshu University, Matsumoto 390-8621, Japan\bigskip\\
}

\date{October 16, 2008}

\maketitle

\begin{abstract}
\noindent
We study the five-dimensional $SU(3)_c\times U(1)_C$ gauge theory on the orbifold $S^1/Z_2$ with a mixed Chern-Simons term. 
We particularly pay attention to the realization of the dynamical rearrangement of the theta parameter for $SU(3)_c$.
It is shown that the physics remains invariant under a large gauge transformation which even changes the action, completely removing the theta parameter, because of the Hosotani mechanism for the $U(1)_C$ gauge interactions.
In other words, we show that the theta parameter can be regarded as a boundary condition for the orbifolding in light of the dynamical rearrangement.
\end{abstract}

\newpage

\section{Introduction}

When we consider a quantum field theory in a non simply connected space $\mathcal{X}$, the fields are first defined in the universal cover of $\mathcal{X}$, say $\mathcal{D}$, and then the ones living in $\mathcal{X}$ are obtained by the identification of the fields at different points of $\mathcal{D}$ which are identified to yield $\mathcal{X}$.
The field living in $\mathcal{D}$ ($\mathcal{X}$) is sometimes called the one in the upstairs (downstairs) picture.
The identification of the fields can be up to a (gauge) transformation which leave the action invariant.
That is, the values of the fields at two points in $\mathcal{D}$ that project to the same point in $\mathcal{X}$ need not be identical, but might differ by a gauge transformation.
Especially when the gauge transformation is global, the identification of the whole region of $\mathcal{X}$ is determined by the identifications at the boundary of $\mathcal{X}$.
The identification at the boundary is called the Boundary Conditions (BCs) for the (downstairs) field.\footnote{In the following, we sometimes use the terminology BC also for more general identification.}
In general, different choice of BCs yields different physical systems.

However, different choices of BCs may be physically equivalent under a large gauge transformation that also takes into account the Wilson line phases for the gauge fields related to the BCs.
Furthermore, the resultant gauge symmetry breaking pattern can be different from the one naively expected from the BCs, due to the Wilson line phases whose values are determined by the effective potential generated by the quantum corrections.
This phenomenon is called the dynamical rearrangement and the symmetry breaking/restoration due to the dynamical rearrangement is called the Hosotani mechanism~\cite{Hosotani}.
So far, the dynamical rearrangement is verified for the gauge theories with $\mathcal{X}$ ($\mathcal{D}$) 
being $T^n$ ($R^n$)~\cite{Hosotani}, $S^1/Z_2$ ($R$)~\cite{HHHK,HHK}, 
and $S^1/Z_2$ ($R$) with the non-factorizable warped metric of the Randall-Sundrum type~\cite{OW}.



Choi has proposed a beautiful mechanism to solve the strong CP problem~\cite{Choi} in terms of the five-dimensional $SU(3)_c\times U(1)_C$ gauge theory on the orbifold $S^1/Z_2$ 
with a mixed Chern-Simons term.
\footnote{
The equivalent model is reformulated in terms of the downstairs language~\cite{G&W}.
}$^,$\footnote{
For other attempts to solve the strong CP problem using Chern-Simons terms, see Ref.~\cite{CP-CS}.
}
In the model, the fifth component of the $U(1)_C$ gauge boson becomes the axion and its potential is induced by the QCD instanton effects.
For the Choi's model to work, it is essential that the bulk fields are completely neutral under the $U(1)_C$ so that the effective potential for the $U(1)_C$ Wilson line phase is never generated through the Hosotani mechanism with the bulk field loops. Otherwise, the generated potential, whose scale would be of the order of the inverse compactification radius $R^{-1}$, would dominate over the QCD instanton potential, spoiling the whole idea.

In this paper, we will show that a large gauge transformation can actually erase the QCD theta term that is placed at a boundary of the $S^1/Z_2$.
Precisely speaking, this is not a symmetry of the action because the Chern-Simons term is not invariant under this transformation.
(That is why it can erase the theta term at the boundary.)
However, we will show that the physics remains intact under this transformation even though it is not a symmetry of the action.
To show the invariance taking into account the quantum corrections, we do put a bulk complex scalar field 
that is charged under the $U(1)_C$ into the Choi's setup, 
though this is phenomenologically unrealistic as is described above.

As is stated, the mixed Chern-Simons term is invariant only under the transformations whose gauge functions vanish at a boundary (or an infinity) of space-time.
We investigate, for the first time, whether the dynamical rearrangement of the mixed Chern-Simons term and the theta parameter occurs or not under the singular gauge transformation whose gauge function does not vanish at the boundary and is not a symmetry of the action.
We find that the effective potential for the Wilson line phase of the $U(1)_C$ gauge field, generated through one loop quantum corrections, always chooses the physically equivalent value of the strong CP violation.


This paper is organized as follows. 
In Sec. 2, we review the dynamical rearrangement of physical system and extend it to the case involving the field redefinition rather than the symmetry transformation of the action.
In Sec. 3, we study the dynamical rearrangement of the theta parameter
using a five-dimensional gauge theory in the presence of a mixed Chern-Simons term.
Section 4 is devoted to conclusions and discussions.

\section{Dynamical rearrangement of physical system}
We review the dynamical rearrangement of physical system in the general
framework of extra-dimensional gauge theory and extend it so that it involves the field redefinition rather than the symmetry transformation of the action.

\subsection{Boundary conditions}
We consider a gauge theory defined on a product of four-dimensional Minkowski space $M^4$ and an extra non-simply connected space $\mathcal{X}$ and write its universal cover $\mathcal{D}$.
The coordinates for $M^4$ and $\mathcal{D}$ are denoted by $x$ and $y$, respectively.
We assume that $\mathcal{X}$ is 
defined by the following identifications in the covering space
\begin{eqnarray}
y \sim f_l(y),
\label{id}
\end{eqnarray}
where $l$ stands for a label when there are several points identified together.
Let a bulk field $\Phi(x, y)$ be a multiplet of some transformation group $G$
and the Lagrangian density $\mathcal{L}$ be invariant under the $G$-transformation 
$\Phi(x,y)\to\Phi'(x, y)=T_{\Phi}\Phi(x, y)$, with $T_{\Phi}$ being a representation of $G$ on $\Phi$:
\begin{align}
	\mathcal{L}\!\left(\Phi(x,y)\right)
		&=	\mathcal{L}\!\left(\Phi'(x,y)\right).
			\label{L-inv}
  \end{align}
When we require $\mathcal{L}$ to be single-valued on $M^4 \times\mathcal{X}$, i.e.,
\begin{align}
	\mathcal{L}\!\left(\Phi(x,y)\right)
		&=	\mathcal{L}\!\left(\Phi\!\left(x,f_l(y)\right)\right) ,
\label{L-single}
  \end{align}
the field can be identified up to a gauge transformation
\begin{eqnarray}
\Phi(x, f_l(y)) = 
	U_l(x,y)\Phi(x, y),
\label{BC}
\end{eqnarray}
where 
$U_l$ is a representation 
of $G$ on $\Phi$. 

After a large gauge transformation\footnote{This gauge transformation, defined in the universal cover $\mathcal{D}$, is large in the downstairs picture since it is not single-valued in $\mathcal{X}$ and cannot be disentangled to the identity there.}
\begin{eqnarray}
	\Phi(x,y)	\to	\Phi'(x, y) = 
		\Omega(x,y)\Phi(x, y) ,
\label{gauge-tr}
\end{eqnarray}
the new field $\Phi'(x, y)$ satisfies the following identification:
\begin{eqnarray}
\Phi'(x, f_l(y)) = U'_l(x,y)\Phi'(x, y),
\label{newBC}
\end{eqnarray}
with the new BC
\begin{eqnarray}
U'_l(x,y) = \Omega(x, f_l(y))\,U_l(x,y)\,\Omega^{-1}(x, y).
\label{newBC}
\end{eqnarray}
The BC is changed by a large gauge transformation as $U_l \to U'_l (\ne U_l)$, 
but physics (physical symmetries, parameters and spectrum) should be invariant under the gauge transformation.
If this is a right statement, rearrangement of symmetries, parameters and spectrum must occur
after taking a new BC connected to by the large gauge transformation.
This phenomenon is understood by the Hosotani mechanism.

\subsection{Conjugate boundary condition}

We show that there can be another choice of boundary conditions, which we call the conjugate identification.\footnote{This is a special case of the orbifold breaking by outer automorphisms~\cite{Hebecker:2001jb}.}
Instead of the identification~\eqref{BC}, we may think of the following boundary condition
\begin{eqnarray}
	\Phi(x, f_l(y)) = U_l^*(x,y)\Phi^*(x, y) ,
	\label{conj_BC}
\end{eqnarray}
where the asterisk indicates the complex conjugation.\footnote{In Ref.~\cite{Grzadkowski:2005rz}, field identifications that mix particles and anti-particles are utilized.} 
After the gauge transformation~\eqref{gauge-tr},
the new field $\Phi'(x, y)$ now satisfies the following identification,
\begin{eqnarray}
\Phi'(x, f_l(y)) = {U_l'}^*(x,y)\Phi'^*(x, y) ,
\label{newBC}
\end{eqnarray}
where the new BC $U'_l$ is given by
\begin{eqnarray}
U'_l(x,y) = \Omega^*(x, f_l(y))\,U_l(x,y)\,\Omega^{-1}(x, y) .
\label{conj_newBC}
\end{eqnarray}

Now we explain two kinds of formulations of a $U(1)$ gauge theory on $S^1/Z_2$ due to the difference of BCs.
In Choi's model~\cite{Choi}, it was required that the fifth component of the $U(1)_C$ gauge field is $Z_2$ even
\begin{align}
	C_5(x,-y)	&=	+C_5(x,y),
  \end{align}
in order to let its zero mode survive and be identified with the axion.
Then, as a consequence of the normal boundary condition~\eqref{BC}, the gauge coupling had to be $Z_2$ odd since the $y$-derivative must be $Z_2$ odd. The resultant covariant derivative and gauge transformation are
\begin{align}
	D_M\Phi	&=		\left(\partial_M+i\q\epsilon(y)C_M\right)\Phi,	&
	\Phi	&\to	e^{-i\q\epsilon(y)\Lambda}\Phi,
  \end{align}
where $\epsilon(y)=\pm1$ is the $Z_2$ odd step function (for $\pm y>0$ and $|y|<L$, with $L$ being the compactification length).

When we utilize the conjugate boundary condition~\eqref{conj_BC}, the $Z_2$ minus sign can be coming from the complex conjugation. In such a case, we can write down the covariant derivative and the gauge transformation with a normal $Z_2$ even gauge coupling
\begin{align}
	D_M\Phi	&=		\left(\partial_M+i\q C_M\right)\Phi,	&
	\Phi	&\to	e^{-i\q \Lambda}\Phi.
					\label{normalone}
  \end{align}
Since there are some subtleties from the fact that $\epsilon'(y)\propto\delta(y)$,\footnote{See e.g.\ Ref.~\cite{OSY} for a possible regularization.} we will concentrate in our analysis on the latter possibility~\eqref{normalone} that does not involve the $Z_2$ odd gauge coupling.

\subsection{Hosotani mechanism}
For our setup, there are non-trivial Wilson line phases, which are the phases of the eigenvalues of the path ordered integral of the extra dimensional components of the gauge field~$A_y$ along a non-contractable loop $C$ in $\mathcal{X}$:
\begin{align}
	W	&=	P\exp\left(ig\int_C A_y dy\right).
  \end{align}
They cannot be gauged away and are physical degrees of freedom.
The so-called Hosotani mechanism can be summarized as follows.
\begin{itemize}
      \item  The physical vacuum is given by the configuration of Wilson line phases that minimizes the effective potential $V_\text{eff}$.
	\item	The physical symmetries, parameters and spectrum are determined by the combination of the BCs and the expectation value of Wilson line phases.
	\item	Two physical systems are equivalent if they are connected by a large gauge transformation, which is a symmetry of the Lagrangian
\begin{eqnarray}
\Bigl.\mathcal{L} \left(\Phi(x,y)\right)\Bigr|_{\left(\langle A_y \rangle, U_l\right)} 
= \Bigl.\mathcal{L}(\Phi'(x,y))\Bigr|_{\left(\langle A'_y \rangle, U'_l\right)} 
\label{L-gaugeinv}
\end{eqnarray}
and is respected by the effective potential
\begin{align}
V_\text{eff}\!\left(\langle A_y \rangle, U_l\right) = V_\text{eff}\!\left(\langle A'_y \rangle, U'_l\right).
  \end{align}
\end{itemize}

Now we expect that the above argument can be generalized 
to the case where a new field $\widetilde{\Phi}$ is obtained by the redefinition of fields irrespective of the invariance of $\mathcal{L}$.
Let us consider a class of field redefinitions that leaves the S-matrices invariant\footnote{Arbitrary changes of variables are not always allowed~\cite{CWZ}. See also~\cite{KOS} for a theorem related to the independence of the S-matrices from a choice of variables.}
\begin{eqnarray}
\Phi(x, y) \to \widetilde{\Phi}(x, y) = \widetilde{\Phi}\left[\Phi(x,y)\right] .
\label{Phi-redef}
\end{eqnarray}
After the redefinition, the Lagrangian $\mathcal{L}$ changes its form into $\widetilde{\mathcal{L}}$:
\begin{eqnarray}
\widetilde{\mathcal{L}}(\widetilde{\Phi}(x,y)) = \mathcal{L}(\Phi(x,y))  .
\label{L-redef}
\end{eqnarray}
The BC in general can become different $U_l \to \widetilde{U}_l (\ne U_l)$ by the redefinition of fields.
Now $\widetilde{U}_l$ is not necessarily a linear transformation like Eq.~\eqref{BC}
but a transformation defined by
\begin{eqnarray}
\widetilde{\Phi}\left(x, f_l(y)\right) = \widetilde{\Phi}\left[ {\Phi}\left(x, f_l(y)\right) \right]
= \widetilde{\Phi}\left[U_l(x,y)\Phi(x, y)\right] \equiv \widetilde{U}_l(\widetilde{\Phi}(x, y)) .
\label{widetildeBC}
\end{eqnarray}

Here we assert that physics should be same after the redefinition of fields.
In the same way as~\eqref{L-gaugeinv}, 
we expect that the invariance of the physical symmetries, parameters and spectrum 
can be guaranteed by the transformation law of $\mathcal{L}$
\begin{eqnarray}
\Bigl.\mathcal{L}(\Phi(x,y))\Bigr|_{\left(\langle A_y \rangle, U_l\right)} 
= \left.\widetilde{\mathcal{L}}(\widetilde{\Phi}(x,y))\right|_{\left(\langle \widetilde{A}_y \rangle, \widetilde{U}_l\right)} 
\label{L-redefinv}
\end{eqnarray}
by using the relation
$V_\text{eff}(\langle A_y \rangle, U_l) = \widetilde{V}_\text{eff}(\langle \widetilde{A}_y \rangle, \widetilde{U}_l)$.

\section{5D gauge theory with mixed Chern-Simons term}
The mixed Chern-Simons term is a CP non-conserving term and deeply related to the strong CP problem. 

\subsection{Strong CP problem}
First we briefly review the strong CP problem.
The strong CP problem is a naturalness problem that asks why the CP-violating phase in QCD is extremely small~\cite{strongCP}.
The non-observation of the neutron electric dipole moment suggests $|\bar{\theta}| \le O(10^{-10})$.
The parameter $\bar{\theta}$ is a physical one unless there is an exact global symmetry that can make $\bar{\theta}$ to be zero,
in which case the value of $\bar{\theta}$ is determined dynamically
by introducing a corresponding physical degrees of freedom.

Three possible solutions have been proposed to solve the strong CP problem.
First one is that one of quarks is massless and then $\bar{\theta}$ is made to be zero by the chiral transformation. This possibility seems to be ruled out by experiment.
Second one is the so-called Peccei-Quinn mechanism~\cite{PQ} involving a light pseudo Nambu-Goldstone boson called axion~\cite{axion}.
In the model, the Peccei-Quinn symmetry $U(1)_\text{PQ}$ couples to the QCD anomaly and $\bar{\theta}$ is made to be zero dynamically by the potential generated by the QCD instanton effects.
Third one is that the CP transformation is an exact symmetry in an underlying high-energy theory, 
and it is broken very weakly in the low-energy theory~\cite{exactCP}.

The Peccei-Quinn mechanism is most popular, but there are two theoretical problems.
One is how to suppress contributions from other explicit $U(1)_\text{PQ}$ breaking terms
such as higher-dimensional operators induced by the possible quantum gravity effects.
The other is how to get the axion decay constant $f_a$ naturally within the narrow window $f_a = 10^{10 \sim 12}$GeV, where the constraint on $f_a$ originates from astrophysical and cosmological bounds.

\subsection{$U(1)_C$ model}

We consider an extension of the standard model (SM)
by introducing an extra gauge symmetry $U(1)_C$ in an extra dimension.
The theory is defined on $B \equiv M^4 \times (S^1/Z_2)$ where $S^1/Z_2$ is the one-dimensional orbifold.
We separate the coordinates $x^M$ ($M=0,1,2,3,5$) into the uncompactified four-dimensional ones $x^\mu$ (or $x$) and the compactified one $y$.
The $S^1/Z_2$ is obtained by dividing the circle $S^1$ (with the identification $y \sim y + 2\pi R$)
by the $Z_2$ transformation $y \to -y$, so that the point $y$ is identified with $-y$.
Then the $S^1/Z_2$ is regarded as an interval with length $\pi R$, with $R$ being the $S^1$ radius.
Both end points $y = 0$ and $\pi R$ are fixed points under the $Z_2$ transformation.
All the SM field except for the gluon are assumed to be localized at the fixed point $y = 0$, 
on the basis of the brane world scenario.\footnote{
For simplicity, we concentrate on the quark and QCD sector neglecting the leptons and electroweak interactions hereafter. It is straightforward to incorporate them.} 

There exist two kinds of four-dimensional fields in our low-energy theory.
One is the brane field which lives only at the boundary, 
and the other is the zero mode stemming from the five-dimensional bulk field.
The Kaluza-Klein (KK) modes do not appear in our low-energy world because they have heavy masses of order $1/R$, as far as $R$ is small enough $1/R\gtrsim\text{TeV}$.
For simplicity, we have assumed that quarks fields $\psi_f$ are localized on the $y=0$ brane.
We assume that the $SU(3)_c$ gauge bosons $A_M^a$ and an extra $U(1)$ gauge boson $C_M$ are bulk fields with $Z_2$ even parities and $Z_2$ odd parities, respectively.

As is discussed in Introduction,
we also introduce a $U(1)_C$ charged bulk complex scalar field $\Phi$ into the Choi's model~\cite{Choi}.
Its real and imaginary parts form a doublet under the $Z_2$ transformation
and generate the effective potential for the $U(1)_C$ Wilson line phase.
Let the bulk fields satisfy the following BCs
\begin{align}
A_{\mu}^a(x, -y) &=  A_{\mu}^a(x, y) , & 
A_{\mu}^a(x, y + 2 \pi R) &= A_{\mu}^a(x, y) , \nonumber\\
A_{5}^a(x, -y) &=  - A_{5}^a(x, y) , & 
A_{5}^a(x, y + 2 \pi R) &= A_{5}^a(x, y) , \nonumber\\
C_{\mu}(x, -y) &=  - C_{\mu}(x, y) , &
C_{\mu}(x, y + 2 \pi R) &= C_{\mu}(x, y) , \nonumber\\
C_{5}(x, -y) &=   C_{5}(x, y) , & 
C_{5}(x, y + 2 \pi R) &= C_{5}(x, y) , \nonumber\\
\Phi(x, -y) &=  \Phi^*(x,y), & 
\Phi(x,y+2\pi R) &= e^{2\pi i\beta}\Phi(x,y), 
\label{BCs}
\end{align}
%
where $\beta$ is a twist parameter.
Several remarks are in order.
The gauge symmetry $U(1)_C$ is explicitly broken by the above BCs at the compactification scale $1/R$.
From the above BC (\ref{BCs}), the $C_5(x,y)$ and $\Phi(x,y)$ are expanded as
\begin{align}
	C_5(x,y)
		&=	{1\over\sqrt{2\pi R}}C_5^{(0)}(x)	
			+{1\over\sqrt{\pi R}}\sum_{n=1}^\infty C_5^{(n)}(x)\cos{n\pi y\over R} ,\\
      \Phi(x,y)
		&=	{1 \over 2\sqrt{\pi R}}\sum_{n=-\infty}^\infty
			\phi_n(x)\exp{i(n+\beta)y\over R}
			\label{Phi-expansion} ,
\end{align}
where $\phi_n(x)$s are four-dimensional real scalar fields. 
Zero modes reside in $A_\mu$ and $C_5$ and that $\Phi$ also yields a one if $\beta=0$.
We will see in the following that the boundary condition $\beta$ and the Wilson line for $C_5$ are related.

Under the $U(1)_C$ gauge transformation, $C_M(x,y)$ and $\Phi(x,y)$ transform as 
\begin{align}
\hspace{-1cm} C_M(x, y) &\to C_M(x,y) - \partial_M \Lambda(x,y) , & 
\Phi(x, y) &\to e^{i\q \Lambda(x,y)} \Phi(x,y) ,
\label{U1C}
\end{align}
where $\q $ is the $U(1)_C$ charge of $\Phi$ with mass dimension $[\q ]=-1/2$,
and $\Lambda(x,y)$ is a gauge transformation function.
We assume that $\Phi$ is neutral under $SU(3)_c$ 
and take the covariant derivative of $\Phi$ as
\begin{eqnarray} 
D_M\Phi(x, y) \equiv  \left(\partial_M + i \q  C_M(x, y)\right)\Phi(x, y) .
\label{D}
\end{eqnarray}
We can check that the covariant derivative $D_M\Phi(x, y)$ satisfies the following BCs
\begin{align}
\left(\partial_\mu + i \q  C_\mu(x, -y)\right)\Phi(x, -y) 
& = \left(\left(\partial_\mu  + i \q  C_\mu(x, y)\right)\Phi(x, y)\right)^* ,
\label{BCs-DPhimu}\\
\left(\partial_{-y} + i \q  C_5(x, -y)\right)\Phi(x, -y) 
& = - \left(\left(\partial_y  + i \q  C_5(x, y)\right)\Phi(x, y)\right)^* .
\label{BCs-DPhiy}
\end{align}

We require that the action should possess gauge invariance consistent with the BCs.
Then the action is written as
\begin{eqnarray}
S_{\rm 5D}= \int d^4x \int dy \left[\mathcal{L}_\text{bulk}
+ \mathcal{L}_\text{brane} \delta(y) + \mathcal{L}'_\text{brane}\delta(y-\pi R)\right] ,
\label{5Daction}
\end{eqnarray}
where $\mathcal{L}_\text{bulk}$ and $\mathcal{L}_\text{brane}$ 
are the related bulk and brane Lagrangian densities, respectively, given by
\begin{eqnarray}
&~& \mathcal{L}_\text{bulk} = -\frac{1}{2} \tr(F_{MN} F^{MN}) 
 - \frac{1}{4}\left(\partial_M C_N - \partial_N C_M\right)\left(\partial^M C^N - \partial^N C^M\right) 
\nonumber \\
&~& ~~~~~~~~ + |D_M \Phi|^2 +  {\kappa\over 5!} \varepsilon^{MNLOP}C_M \tr\left(F_{NL}F_{OP}\right) , 
\label{LBulk}\\
&~& \mathcal{L}_\text{brane} = \overline{\psi}_f i \gamma_{\mu} D^{\mu} \psi_f 
 + \frac{\theta}{32\pi^2} \varepsilon^{\mu\nu\rho\sigma}\tr\left(F_{\mu\nu}^{(0)} F_{\rho\sigma}^{(0)}\right)  ,
\label{Lbrane}
\end{eqnarray}
where 
the fourth term in $\mathcal{L}_\text{bulk}$ is the mixed Chern-Simons term with $\kappa$ being the coupling constant of mass dimension $[\kappa]=-1/2$ and the second term in $\mathcal{L}_\text{brane}$ is the theta term with
$\theta$ being the QCD vacuum angle on the brane at $y=0$.\footnote{
After the breakdown of $SU(2)_L \times U(1)_Y$ and the re-definition of quark fields' phase,
the parameter $\theta$ becomes the effective one $\bar{\theta} \equiv \theta + {\mbox{argdet}}(M_u M_d)$
where $M_{u,d}$ are mass matrices of the up and down-type quarks.}
We do not specify the brane Lagrangian density $\mathcal{L}'_\text{brane}$ at $y=\pi R$ for simplicity.
We note that we have put the theta term only for the zero mode $F_{\mu\nu}^{(0)}$ 
though there can be other terms 
$\displaystyle{\theta_n \varepsilon^{\mu\nu\rho\sigma}\tr\left(F_{\mu\nu}^{(n)} F_{\rho\sigma}^{(n)}\right)}$, 
$\displaystyle{\Theta \varepsilon^{\mu\nu\rho\sigma}\tr\left(F_{\mu\nu} F_{\rho\sigma}\right)}$, etc.

The Kaluza-Klein mass of $\Phi(x,y)$ is obtained by using the mode expansion (\ref{Phi-expansion})
\begin{eqnarray}
\int_{-\pi R}^{\pi R} dy |(\partial_y + i q_{\Phi} \langle C_5 \rangle)\Phi(x,y)|^2 
= \sum_{n=-\infty}^{\infty} \frac{(n+\beta+\gamma)^2}{R^2} (\phi_n(x))^2 ,
\label{kineticPhi}
\end{eqnarray}
where
\begin{align}
	\gamma
		&\equiv	
				\q  R\langle C_5\rangle
		=		\q \sqrt{R\over2\pi}\langle C_5^{(0)}\rangle. 
\end{align}
The $\gamma$ can be regarded as the expectation value of Wilson line phase multiplied by $q_{\Phi}/(2\pi)$ such as 
$\displaystyle{\gamma =	\frac{\q }{2\pi}\int_{-\pi R}^{\pi R} \langle C_5 \rangle dy}$.

Under the gauge transformation $\delta C = d\Lambda$, 
the mixed Chern-Simons term in the action transforms as
\begin{align}
	\delta \left(\kappa\int_B C\tr (FF)\right)
		=	\kappa\int_B d\Lambda \tr (FF) 
		=	\kappa\int_B d\left(\Lambda \tr (FF)\right)
		=	\kappa\int_{\partial B} \Lambda \tr (FF) ,	\label{mCS-trans}
  \end{align}
where the wedge product is omitted and $\partial B$ is the boundary of $B$ (in the downstairs picture).
Therefore the mixed Chern-Simons term is invariant for the gauge transformations $\Lambda(x,y)$ that vanish at $\partial B$.

\section{Dynamical rearrangement}

In our model, the $C_5$ couples to the QCD anomaly $\varepsilon^{\mu\nu\rho\sigma}F_{\mu\nu}F_{\rho\sigma}$ in the mixed Chern-Simons term
and so one may expect that it can play a role of the axion~\cite{Choi}.
As Choi has suggested, all bulk matter fields are required to be neutral in order the Peccei-Quinn mechanism to work well.
In fact, in our particle contents with the bulk charged field $\Phi$, the expectation value of $C_5$ is determined by the Hosotani mechanism and we will see that the Peccei-Quinn mechanism does not work well without an extreme fine-tuning among parameters.
Our model with $\Phi$ is not intended to provide a phenomenological application nor to solve the strong CP problem, but to study the dynamical rearrangement of physical system with the topological term.
We shall apply the Hosotani mechanism to our model.

\subsection{Before large transformation}
 
The one-loop effective potential for the background configuration of $U(1)$ gauge boson, $C_M^\text{bg}$, is 
derived through the contribution from $\Phi$ and calculated as~\cite{KLY,HHHK}
\begin{align}
V^\text{4D}_\text{eff}[\gamma]
&= -{1\over2}\int \frac{d^4p_E}{(2\pi)^4} \sum_{n=-\infty}^{n=\infty} 
 \ln\left(p_E^2 + \left(\frac{n+\beta+\gamma}{R}\right)^2\right) 
\nonumber \\
&= E_0 + \frac{3}{64\pi^6 R^4}\sum_{n=1}^{\infty} \frac{1}{n^5}\left(1-\cos 2\pi n(\beta+\gamma)\right) ,	\label{Veff}
\end{align}
where $p_E$ is a four-dimensional Euclidean momentum
and $E_0$ is a divergent but $\gamma$-independent constant.
We find that the physical vacuum is realized at $\beta + \gamma=0$ (or $\pi$)
and then the theta term on our brane turns out to be
\begin{align}
\mathcal{L}^\theta_\text{brane}
	&=	\left(\frac{\theta}{32\pi^2}+{\kappa\over4!}\langle C_5\rangle\right)\varepsilon^{\mu\nu\rho\sigma}\tr\left(F^{(0)}_{\mu\nu} F^{(0)}_{\rho\sigma}\right)
		+{\kappa\over4!}\langle C_5\rangle \varepsilon^{\mu\nu\rho\sigma}\sum_{n=1}^\infty\tr\left(F^{(n)}_{\mu\nu} F^{(n)}_{\rho\sigma}\right)
\nonumber \\
	&=	\left(\frac{\theta}{32\pi^2} - \frac{\kappa}{4!\q  R}\beta\right) \varepsilon^{\mu\nu\rho\sigma}\tr\left(F^{(0)}_{\mu\nu} F^{(0)}_{\rho\sigma}\right)
		- \frac{\kappa}{4!\q  R}\beta  \varepsilon^{\mu\nu\rho\sigma} \sum_{n=1}^\infty\tr\left(F^{(n)}_{\mu\nu} F^{(n)}_{\rho\sigma}\right),
\label{theta-term}
\end{align}
where we have chosen the vacuum value $\beta + \gamma=0$. 

The action related to $C_5^{(0)}(x)$ is given by
\begin{align}
S_{C_5^{(0)}}
	&=	\int d^4x\,\frac{1}{2} \left(\partial_\mu C_5^{(0)} \partial^{\mu} C_5^{(0)} + m_C^2 {C_5^{(0)}}^2\right) \nonumber \\
	&\quad
		+ \int d^4x \int dy \left({\kappa\over4!} \varepsilon^{5\mu\nu\rho\sigma}C_5 \tr\left(F_{\mu\nu}F_{\rho\sigma}\right) 
		+ \left|\left(\partial_y + i\q  C_5\right) \Phi\right|^2 \right),
\label{S-C5}
\end{align}
where $m_C$ is the mass of $C_5^{(0)}$ obtained from $V_\text{eff}^\text{4D}$ as
\begin{align}
m^2_C
	&=	\left.{\partial^2 V_\text{eff}^\text{4D}\over {\partial C_5^{(0)}}^2}\right|_{\beta+\gamma=0}
	 =	\frac{3\q ^2}{32\pi^5 R^3} \zeta(3) ,
\label{mC}
\end{align}
where $\zeta$ is the Riemann zeta function.
As the $C_5^{(0)}$ also acquires heavy mass of $O(1/R)$, $C_M$ completely disappear from the low-energy spectrum.

Now let us confirm that the Peccei-Quinn mechanism does not work in the presence of $V_\text{eff}^\text{4D}[\gamma]$.
By the QCD instanton effect, the following potential is induced
\begin{eqnarray}
V_\text{PQ}\!\left[C_5^{(0)}\right] 
= m_{\pi}^2 f_{\pi}^2 \left(1 - \cos\!\left(\bar{\theta}+{4\pi^2\kappa\over3\sqrt{2\pi R}}C_5^{(0)}\right)\right) ,
\label{VeffPQ}
\end{eqnarray}
where $m_{\pi}$ and $f_{\pi}$ are the pion mass and the pion decay constant, respectively.
Since the KK mass scale must be much larger than the pion mass $1/R \gg m_{\pi}$,
the minimum of the total effective potential $V_{\rm tot} = V_\text{eff}^\text{4D}[\gamma] + V_\text{PQ}[C_5^{(0)}]$ 
is still located at $\beta + \gamma = 0$ (or $\pi$) up to the correction of order $O(m_{\pi}^2 f_{\pi}^2 R^4{\bar{\theta}})$.
Hence the effective theta parameter $\theta_\text{eff}\equiv\bar{\theta}+{4\pi^2\kappa\over3\sqrt{2\pi R}}\langle C_5^{(0)}\rangle$ does not vanish unless there is a fine-tuning between the parameters $\bar\theta$ and $\beta$.
The theta term can be erased when suitable BCs are taken, but it is nothing but a fine-tuning.

\subsection{After large transformation}

As shown in (\ref{mCS-trans}), 
the mixed Chern-Simons term is gauge invariant if the $\Lambda(x,y)$ vanishes at $\partial B$.
By contrast, we can change the value of $\theta$ in $\mathcal{L}_\text{brane}$ 
by taking a specific $\Lambda(x,y)$ which does not vanish at the boundary $y = 0$.
That is, the theta term is absorbed into the mixed Chern-Simons term by a field redefinition, $\widetilde{C} = C + d\Lambda$
with a suitable $\Lambda(x,y)$.
We shall examine whether dynamical rearrangement of the theta parameter occurs or not in this situation.

We use the following gauge transformation function,\footnote{Note that $\Lambda(x,y)$ is not periodic in~$y$ but a monotonically increasing function. Though the expression~\eqref{vartheta} appears to be discontinuous, it can be obtained by taking the derivative of a continuous function. This seemingly discontinuous function and its derivative (being delta functions) are essential to show the very existence of the solution to the Einstein equation that leads to the Randall-Sundrum geometry~\cite{Randall:1999ee}. One may instead take the smoothed form, say, around $y=0$: $\epsilon(y)=\tanh(y/\delta)$ with $\delta$ being an infinitesimal~\cite{OW}.}
\begin{align}
	\Lambda(x,y)
		&=	-{3(2n+1)R\theta\over4\pi\kappa} \quad
			\text{for $2n\pi R<y<2(n+1)\pi R$ with $n\in \mathbb{Z},$}
			\label{vartheta}\\
	\partial_y \Lambda(x,y)
		&=	-\frac{3R\theta}{2\pi\kappa} \sum_{n=-\infty}^\infty \delta(y-2n\pi R)
			\label{dvartheta} ,
\end{align}
defined on the whole covering space $-\infty<y<\infty$.
After the gauge transformation or the field redefinition, 
the theta term at the $y=0$ brane is absorbed into the mixed Chern-Simons term and the Lagrangian density is written as
\begin{align}
	\widetilde{\mathcal{L}}_\text{bulk}
		&=	-\frac{1}{2} \tr\left(F_{MN}F^{MN}\right) 
			- \frac{1}{4}\left(\partial_M \widetilde{C}_N - \partial_N \widetilde{C}_M\right)\left(\partial^M \widetilde{C}^N - \partial^N \widetilde{C}^M\right) 
\nonumber \\
		&\quad
			+ \left|(\partial_M + i \q  \widetilde{C}_M) \widetilde{\Phi}\right|^2 +  {\kappa\over5!} \varepsilon^{MNLOP} \widetilde{C}_M \tr\left(F_{NL} F_{OP}\right) , 
\label{LBulktilde}\\
	\widetilde{\mathcal{L}}_\text{brane}
		&=	\overline{\psi}_f i \gamma_{\mu} D^{\mu} \psi_f
			-{\theta\over32\pi^2}\varepsilon^{\mu\nu\rho\sigma}\sum_{n=1}^\infty \tr\left(F^{(n)}_{\mu\nu}F^{(n)}_{\rho\sigma} \right),
\label{Lbranetilde}
\end{align}
where the new fields $\widetilde{C}_M$ and $\widetilde{\Phi}$ are given by
\begin{align}
	\widetilde{C}_M(x, y)
		&=	C_M(x,y) - \partial_M \Lambda(x,y) , & 
	\widetilde{\Phi}(x, y)
		&=	e^{i\q  \Lambda(x,y)} \Phi(x,y) .
\label{tildeC}
\end{align}
We note that the large gauge transformation changes the brane action.
The BCs for $\widetilde{C}_M$ and $\widetilde{\Phi}$ are given by
\begin{align}
	\widetilde{C}_{\mu}(x, -y)
		&=	- \widetilde{C}_{\mu}(x, y) , &
	\widetilde{C}_{\mu}(x, y + 2 \pi R)
		&=	\widetilde{C}_{\mu}(x, y) ,\nonumber\\
	\widetilde{C}_{5}(x, -y)
		&=	\widetilde{C}_{5}(x, y) ,	&
	\widetilde{C}_{5}(x, y + 2 \pi R)
		&=	\widetilde{C}_{5}(x, y) ,\nonumber\\
	\widetilde{\Phi}(x, -y)
		&=    \widetilde{\Phi}^*(x, y) , &
	\widetilde{\Phi}(x, y + 2 \pi R)
		&=	e^{2\pi i \left(\beta - \frac{3\q R}{4\pi^2 \kappa}\theta\right)} \widetilde{\Phi}(x, y) .
\label{BCs-Phi}
\end{align}
The one-loop effective potential for $\widetilde{C}_M^\text{bg}$, is 
derived through the contribution from $\widetilde{\Phi}$ and calculated as
\begin{align}
	V^\text{4D}_\text{eff}[\widetilde{\gamma}]
		&=	-{1\over2}\int \frac{d^4p_E}{(2\pi)^4}\sum_{n=-\infty}^{n=\infty}
			\ln\left(p_E^2 + \left(\frac{n+\beta - \frac{3\q R}{4\pi^2 \kappa}\theta+ \widetilde{\gamma}}{R}\right)\right) 
\nonumber \\
		&=	E_0 + \frac{3}{64\pi^6 R^4}\sum_{n=1}^{\infty} \frac{1}{n^5}
\left(1-\cos\!\left[2\pi n\left(\beta - \frac{3\q R}{4\pi^2 \kappa}\theta + \widetilde{\gamma}\right)\right]\right) ,
\label{Veff}
\end{align}
where
\begin{align}
	\widetilde{\gamma}
  \equiv			\q  R\langle \widetilde{C}_5 \rangle
		=		\q \sqrt{R\over2\pi}\langle \widetilde{C}_5^{(0)} \rangle 
            =		\gamma+{3\q  R\over 4\pi^2\kappa}\theta. 
\end{align}
The $\widetilde{\gamma}$ can also be regarded as the expectation value of Wilson line phase multiplied 
by $q_{\Phi}/(2\pi)$ such that
\begin{align}
\widetilde{\gamma} = \frac{\q }{2\pi}\int_{-\pi R}^{\pi R} \langle \widetilde{C}_5 \rangle dy
            =   \frac{\q }{2\pi}\int_{-\pi R}^{\pi R} \langle {C}_5 \rangle dy
                  - \frac{\q }{2\pi}\int_{-\pi R}^{\pi R} \partial_y \Lambda dy
		=		\gamma+{3\q  R\over 4\pi^2\kappa}\theta.
\end{align}
The physical vacuum is realized at $\beta - \frac{3\q R}{4\pi^2 \kappa}\theta + \widetilde{\gamma} = 0$
and then the resultant theta term turns out to be the same as (\ref{theta-term}),
\begin{align}
\widetilde{\mathcal{L}}^\theta_\text{brane}
	&=	{\kappa\over4!}\langle \widetilde C_5\rangle\varepsilon^{\mu\nu\rho\sigma}\tr\left(F^{(0)}_{\mu\nu} F^{(0)}_{\rho\sigma}\right)
		+\left({\kappa\over4!}\langle \widetilde C_5\rangle-{\theta\over32\pi^2}\right) \varepsilon^{\mu\nu\rho\sigma}\sum_{n=1}^\infty\tr\left(F^{(n)}_{\mu\nu} F^{(n)}_{\rho\sigma}\right)
\nonumber \\
	&=	\left(\frac{\theta}{32\pi^2} - \frac{\kappa}{4!\q  R}\beta\right) 
			\varepsilon^{\mu\nu\rho\sigma}\tr\left(F^{(0)}_{\mu\nu} F^{(0)}_{\rho\sigma}\right)
		- \frac{\kappa}{4!\q  R}\beta  \varepsilon^{\mu\nu\rho\sigma} \sum_{n=1}^\infty\tr\left(F^{(n)}_{\mu\nu} F^{(n)}_{\rho\sigma}\right).
\label{theta-term-revisited}
\end{align}
In this way, the dynamical rearrangement of the theta parameter is realized.
We have shown that the five-dimensional 
 gauge theory with a mixed Chern-Simons term possesses
 the ``symmetry'' of the dynamical rearrangement.

\section{Conclusions}

We have investigated the dynamical rearrangement in the higher-dimensional gauge theory on $S^1/Z_2$ with the mixed Chern-Simons term.
In the analysis, the theta term is absorbed into the mixed Chern-Simons term by the special type of large gauge transformation, or more precisely the field redefinition.
Although the existence of the Chern-Simons term breaks the five-dimensional total gauge symmetry, we have generalized the Hosotani mechanism to incorporate the field redefinition (rather than the symmetry of the Lagrangian) and have shown that the theta parameter can be regarded as the boundary condition for the orbifolding in light of the dynamical rearrangement.
We hope that our study would provide better insight and shed light on the higher dimensional solution to the strong CP problem.

\section*{Acknowledgements}
We thank R.\ Kitano for helpful comments.
K.O.\ thanks B.\ Grzadkowski and A.\ Weiler for useful discussions and T.\ Watari for helpful comments.
We thank the Yukawa Institute for Theoretical Physics at Kyoto University, 
where this work was initiated during the YITP workshop $\lq\lq$Summer Institute 2007" (YITP-W-07-12).
This work was supported in part by Scientific Grants from the Ministry of Education and Science, 
Nos.~16540258, 17740146 (N.H.), 18204024, 18540259 (Y.K.), and
19740171 (K.O.).

\end{document}